\documentclass[aps,pra,twocolumn,nofootinbib]{revtex4-1}
\usepackage{epsfig}
\usepackage[colorlinks,linkcolor=blue,anchorcolor=blue,citecolor=blue,urlcolor=blue,breaklinks=true]{hyperref}
\usepackage{graphics}
\usepackage{color}
\usepackage{amsmath}
\usepackage{bm}
\begin{document}
\author{Guo-Hua Liang$^{1}$}
\email{guohua@nju.edu.cn}
\author{Yong-Long Wang$^{2}$}
 \email{wangyonglong@lyu.edu.cn}
\author{Meng-Yun Lai$^{1,2}$}
\author{Hui Liu$^{1}$}
\email{liuhui@nju.edu.cn}
\author{Hong-Shi Zong$^{1,3,4}$}
 \email{zonghs@nju.edu.cn}
\author{Shi-Ning Zhu$^{1}$}
%\email{zhusn@nju.edu.cn}
%\address{$^{1}$ Key Laboratory of Modern Acoustics, MOE, Institute of Acoustics, and Department of Physics, Collaborative Innovation Center of Advanced Microstructures, Nanjing University, Nanjing 210093, China}

\address{$^{1}$ National Laboratory of Solid State Microstructures, School of physics, Collaborative Innovation Center of Advanced Microstructures, Nanjing University, Nanjing 210093, China}
\address{$^{2}$ School of Physics and Electronic Engineering, Linyi University, Linyi 276005, China}
\address{$^{3}$ Joint Center for Particle, Nuclear Physics and Cosmology, Nanjing 210093, China}
\address{$^{4}$ State Key Laboratory of Theoretical Physics, Institute of Theoretical Physics, CAS, Beijing, 100190, China}
%\address{$^{2}$Joint Center for Particle, Nuclear Physics and Cosmology, Nanjing 210093, China}
%\address{$^{3}$State Key Laboratory of Theoretical Physics, Institute of Theoretical Physics, CAS, Beijing, 100190, China}

\title{Pseudo-magnetic field and effective spin-orbit interaction for a spin-1/2 particle confined to a curved surface}
\begin{abstract}
By considering the spin connection, we deduce the effective equation for a spin-1/2 particle confined to a curved surface with the non-relativistic limit and in the thin-layer quantization formalism.
We obtain a pseudo-magnetic field and an effective spin-orbit interaction generated by the spin connection. Geometrically, the pseudo-magnetic field is proportional to the Gaussian curvature and
the effective spin-orbit interaction is determined by the Weingarten curvature tensor. Particularly,
we find that the pseudo-magnetic field and the effective spin-orbit interaction can be employed to
separate the electrons with different spin orientations. All these results are demonstrated in two
examples, a straight cylindrical surface and a bent one.
%\bigskip
%\noindent PACS Numbers: 12.38.Mh, 12.39.-x, 25.75.Nq
\end{abstract}

\pacs{}

\maketitle

\section{INTRODUCTION}
During the last decade, inspired by the technical progress in the fabrication of low-dimensional nanostructures~\cite{201000591,ph500144s,C4NR00330F,201705630,PhysRevLett.119.033902,Liu2013Trapping}, there have been many theoretical and experimental researches on the dynamics in low-dimensional curved spaces, involving condensed matter~\cite{PhysRevLett.113.227205,PhysRevLett.115.256801,nl504305s,135801}, optics~\cite{PhysRevA.78.043821,PhysRevX.4.011038,Bekenstein2017Control}, and magnetism~\cite{PhysRevLett.112.257203,0022-3727-49-36-363001}.
%Generally the geometry of these systems are viewed as a curved two-dimensional (2D) surface embedded in a three-dimensional (3D) Euclidean space, and the particle motion is constrained to the surface by strong confinement implemented by forces normal to the surface.
Due to the appearance of curvature, physical quantities of the particle in the low-dimensional curved space may lose some symmetries associated with space coordinates, such as translational symmetry and rotational symmetry, manifesting novel and abundant properties different from flat cases.
To obtain the curvature effects theoretically, a dimensionally reduced method called the thin-layer procedure (TLP) or confining potential approach was introduced by Jensen, Koppe~\cite{JENSEN1971586}
and da Costa~\cite{PhysRevA.23.1982}. By applying the approach to scalar fields, the well-known geometric potential appears naturally in the effective low-dimensional Hamiltonian. Later, TLP has been applied to electromagnetic fields for two-dimensional (2D)~\cite{PhysRevA.78.043821} and quasi one-dimensional (1D)~\cite{PhysRevA.97.033843} curved geometry. Compared with the case of scalar fields, the connections of vector fields contribute the spin-orbit interaction (SOI) to the effective Hamiltonian as an extra geometric effect, which causes the optical spin-Hall effect (SHE)~\cite{Bliokh2008Geometrodynamics, 10.2307,BLIOKH20151,PhysRevLett.93.083901,Petersen67,xiao0003,PhysRevLett.120.243901}.

%One of the most typical example is the spin-Hall effect (SHE) of light~\cite{Bliokh2008Geometrodynamics,10.2307,BLIOKH20151,PhysRevLett.93.083901,Petersen67}. When a light beam propagates along a curved trajectory because of inhomogeneous refractivity or confinement, the spin and orbital angular momentum are no longer independently conserved, but show their coupling through a polarization-dependent deflection~\cite{Bliokh2008Geometrodynamics}. This spin-orbit interaction (SOI) of photons, which shows the vector field losing translational symmetry on curved manifolds, can be derived from the Maxwell equation consistently~\cite{PhysRevA.46.5199,Bliokh2008Geometrodynamics,PhysRevA.97.033843}.

Certainly, people want to give the explicit SOI Hamiltonian for spinor fields on curved surface. Traditionally, in the planar two-dimensional electron gas (2DEG) systems, there are two main types of SOI, the Rashba SOI and the Dresselhaus SOI, which originate from the inversion asymmetry of the confining potential~\cite{0022-3719-17-33-015} and the lack of bulk inversion asymmetry~\cite{PhysRev.100.580}, respectively. For both of them, the intrinsic or external electric field are considered necessary to generate the SOI. Previous investigations~\cite{PhysRevB.64.085330,PhysRevB.87.174413,PhysRevB.91.245412} have tried to give a theoretical description of the SOI in curved 2DEG systems by writing the electric field induced SOI Hamiltonian in three-dimensional (3D) curvilinear coordinates and using the TLP to get the effective Hamiltonian. These models are capable of describing the part of SOI generated by electric field, but miss the effective SOI induced by the curvature, which is originated from the spin connection.

%However, in two-dimensional electron gas (2DEG) systems, the generation of SOI for electrons is basically rely on the electric field.
%To obtain the curvature effects theoretically, a dimensionally reduced method called the thin-layer procedure or confining potential approach was introduced by Jensen, Koppe~\cite{JENSEN1971586} and da Costa~\cite{PhysRevA.23.1982} in dealing with Schr\"{o}dinger equation.
%This approach is physically more realistic than the intrinsic quantization approach~\cite{RevModPhys.29.377}, and widely employed to give correct descriptions of low-dimensional curved systems. Now it has been developed~\cite{SCHUSTER2003132,PhysRevLett.100.230403} into a versatile method and applied in various areas~\cite{PhysRevA.78.043821,PhysRevA.90.042117,PhysRevLett.112.257203}.
%Among these researches, the SOI in curved low-dimensional structures attracts much attention for its potential applications in spintronics.
% Like photons, we believe an effective SOI of electrons can be derived from the Dirac equation in curved space consistently without any external electric and magnetic field.
 %the confining potential approach~\cite{JENSEN1971586,PhysRevA.23.1982,SCHUSTER2003132,PhysRevLett.100.230403} to get an effective low-dimensional Hamiltonian.

The spin connection, emerging naturally in the covariant derivative of a spinor field, guarantees that a covariant derivative of a spinor transforms still like a spinor. In the language of quantum field theory, it can also be viewed as a non-Abelian gauge field generated by local Lorentz transformations~\cite{16511}. In the 2D curved space, the motion of a fermion should embody the effect of this gauge field~\cite{OUYANG1999297}.Several studies have considered the contribution of the spin connection to the dynamics on curved surfaces~\cite{PhysRevA.90.042117,PhysRevA.48.1861,BRANDT20163036}, however, the explicit effective SOI Hamiltonian from curvature has not be given.
%In Ref. the authors in the effective Pauli equation, but haven't given the explicit effective SOI Hamiltonian.
In addition, it has been found that the SOI from electric fields can also be reformulated in terms of non-Abelian gauge fields~\cite{PhysRevB.73.153306,0305,LEURS2008907,0295}. The coincidence implies that for a spinor field, curvature may in a sense, induce an effective electric field.
Therefore, in this paper, we start from the Dirac equation in a curved space, and try to give a full description about the dynamics of a non-relativistic spin-1/2 particle constrained to a curved surface. %As will be shown, even without intrinsic and external electric fields, an effective SOI can still be generated purely by the geometry.

Another aim of this paper is about the gauge potential from higher dimensions. In Ref.~\cite{SCHUSTER2003132} it emphasizes that a gauge potential is present when the space of states associated with directions normal to the surface is degenerate. In our case the spin degeneracy naturally satisfies the requirement, thus a gauge potential effect is expected. We will show that this effect is a pseudo-magnetic field.
It has been proven theoretically~\cite{Guinea2009Energy,PhysRevB.92.245110} and experimentally~\cite{Levy544} that, a strained graphene sheet can generate a large pseudo-magnetic field, which is explained as a geometry induced gauge field effect in the low-energy effective Dirac equation. We furthermore show that this field exists in curved 2DEG systems.

%Thus we theoretically demonstrate that both the electric and magnetic field induced effects may be reconstructed in correspondingly curved geometries as a result of the spin connection. Our model also gives a glimpse into the symmetry breaking of electroweak SU(2)$\times$U(1) gauge field.

This paper is organized as follows. In Sec. II we briefly give the non-relativistic limit form of Dirac equation in a curved space. In Sec. III, we use the confining potential approach to get an effective equation for a non-relativistic spin-1/2 particle constrained to a curved surface, and analyze the effective SOI and pseudo-magnetic field. In Sec. IV, we explore the effective SOI and pseudo-magnetic field in strait and bent cylindrical surfaces. In Sec. V, we extend the theory to include the model electric and magnetic field. Finally, we present our conclusion in Sec. V.

\section{Non-relativistic limit of Dirac equation in curved space}
In this section, we briefly review the derivation of the non-relativistic limit form of Dirac equation in a curved space and compare it to the ordinary Schr\"{o}dinger equation. Before the derivation, we would like to introduce the meaning of indices used in the following. The Greek letters \{ $\mu,\nu,\dots$\} and \{$\alpha,\beta,\dots$\} run from 0 to 3, denoting curved and flat space-time indices, respectively; the capital Latin letters \{$A,B,\dots$\} and \{$I,J,\cdots$\} run from 1 to 3, denoting curved and flat space indices, respetively, and the lowercases \{$a,b,\dots$\} and \{$i,j,\dots$\} run from 1 to 2, denoting curved and flat 2D space indices, respectively.

In a curved space-time the Dirac equation can be described by~\cite{2009qftc}
\begin{equation}\label{dirac}
(i\hbar \gamma^\mu \nabla_\mu-mc)\Psi=0,
\end{equation}
where $\nabla_{\mu}$ denotes the covariant derivative with $\nabla_\mu=\partial_\mu-\Gamma_\mu$, wherein $\Gamma_\mu$ is the spin connection,
%, and the Greek index $\mu$ run from 0 to 3, denoting the components of the curvilinear coordinates.
and $\gamma^{\mu}$ stand for the gamma matrices in a curved space-time that are expressed
\begin{equation}
\gamma^\mu = E^\mu_\alpha \gamma^\alpha,
\end{equation}
by $\gamma^\alpha$, the usual gamma matrices in a flat space-time. Here
 $E^\mu_\alpha$ are vierbeins with the definition
\begin{equation}
E^\mu_\alpha=\frac{\partial q^\mu}{\partial x^\alpha},
\end{equation}
where $q^{\mu}$ describe the coordinate variables in a curved space-time, $x^{\alpha}$ do those in a flat space-time. By virtue of the vierbeins $E_{\alpha}^{\mu}$, $G^{\mu\nu}$ can be given by
\begin{equation}
G^{\mu\nu}=E^\mu_\alpha E^\nu_\beta \eta^{\alpha\beta},
\end{equation}
where $G^{\mu\nu}$ is the inverse metric tensor in a curved space-time, while $\eta^{\alpha\beta}$ is that in a flat space-time with $\eta^{\alpha\beta}=diag(1,-1,-1,-1)$.
And the spin connection $\Gamma_\mu$ can be expressed as
\begin{equation}
\Gamma_\mu=- \frac{1}{4}\omega_{\mu \alpha\beta} \Sigma^{\alpha\beta},
\end{equation}
where $\Sigma^{\alpha\beta}=[\gamma^\alpha,\gamma^\beta]/2$ are the generators of the Lorentz group in the spinorial space, and
\begin{equation}
\omega_{\mu \alpha\beta}=E_\alpha^\nu (\partial_\mu E_{\nu \beta}-\Gamma_{\mu\nu}^\kappa E_{\kappa \beta}),
\end{equation}
with $\Gamma_{\mu\nu}^\kappa$ being the Christoffel symbol.
%The last term $\mathcal{A}_\mu$ is an arbitrary vector. For convenience, we set $\mathcal{A}_\mu=0$, corresponding to the Fock-Ivanenko spin connection~\cite{Poplawski2009Spacetime}.

For investigating the specially geometric effects, in the present paper we just consider the space curved without time, and we can write $ds^2$ in the following form
\begin{equation}
ds^2=G_{00}dq^0dq^0-G_{AB}dq^A dq^B
\end{equation}
where $G_{00}=1$. As a result, the vierbein $E_{\mu}^{\alpha}$ can be simplified as
\begin{equation}
E_\mu^\alpha=\left(
\begin{array}{ccc}
1 & 0 \\
0& E_A^I
\end{array}
\right).
\end{equation}
%where the flat space indices $I$ run from 1 to 3.
In this case, the time component of spin connection $\Gamma_0$ is naturally vanished, the space components of gamma matrices $\gamma^A$ in Dirac representations are represented as
\begin{equation}
\gamma^A=\left(
\begin{array}{ccc}
0& \sigma^I E_I^A \\
 -\sigma^I E_I^A &0
\end{array}
\right),
\end{equation}
and the space components of the generator $\Sigma^{IJ}$ are
\begin{equation}
\Sigma^{IJ}=\left[
\begin{array}{ccc}
i\epsilon^{IJK}\sigma_K & 0 \\
0 & i\epsilon^{IJK}\sigma_K
\end{array}
\right],
\end{equation}
where $\sigma^I$ are the usual Pauli matrices and $\epsilon^{IJK}$ is the Levi-Civita symbol.

For the convenience of description, we write the wave function $\Psi$ in Eq.~\eqref{dirac} in the following form with two components
\begin{equation}
\Psi=\left[
\begin{array}{ccc}
\chi \\ \phi
\end{array}
\right],
\end{equation}
where $\chi$ and $\phi$ stands for the positive and negative energy solution, respectively.
According to the above discussions, the Dirac equation Eq.~\eqref{dirac} can be rewritten as
\begin{equation}\label{dirac2}
\left[
\begin{array}{ccc}
(i\hbar/c) \partial_t-mc & -\bm{\sigma}\bm{P} \\
 \bm{\sigma}\bm{P} & (-i\hbar/c) \partial_t-mc
\end{array}
\right] \left[
\begin{array}{ccc}
\chi \\ \phi
\end{array}
\right]=0,
\end{equation}
where $\bm{\sigma}\bm{P}=-i\hbar\sigma^A \bar{D}_A=-i\hbar\sigma^K E_K^A(\partial_A+\bar{\Omega}_A)$, wherein $\bar{\Omega}_A$ is the spin connection consisting of two components, that is
\begin{equation}\label{omeg}
\bar{\Omega}_A=\frac{i}{4}\omega_{AIJ} \epsilon^{IJK}\sigma_K.
\end{equation}
From Eq.~\eqref{dirac2}, by eliminating $\phi$, we obtain
\begin{equation}\label{de1}
(E-mc^2)\chi=\bm{\sigma}\bm{P}\frac{1}{E+mc^2}\bm{\sigma}\bm{P}\chi,
\end{equation}
here, $E$ is the total energy.
By defining $E_s=E-mc^2$ and keeping the lowest term in the expansion of $\frac{1}{E+mc^2}$ with $E_s\ll mc^2$, we can obtain the non-relativistic equation
\begin{equation}
E_s \chi=\frac{1}{2m}\bm{\sigma}\bm{P} \bm{\sigma}\bm{P}\chi.
\end{equation}

In a curved space, the covariant derivative of the Dirac matrices is
\begin{equation}
\gamma^B_{\ \ ;A}=\partial_A \gamma^B+\Gamma_{CA}^B \gamma^C-[\Gamma_A,\gamma^B].
\end{equation}
According to the ``tetrad postulate" of van Nieuwenhuizen~\cite{VANNIEUWENHUIZEN1981189}, the curved-space gamma matrices are covariantly constant, which means $\gamma^B_{\ \ ;A}=0$. Accordingly, the Pauli matrices in the curved space satisfying the following equality
\begin{equation}\label{psig}
\partial_A\sigma^B+[\bar{\Omega}_A,\sigma^B]=-\Gamma_{CA}^B \sigma^C,
\end{equation}
and then we can deduce the following expression
\begin{equation}
\begin{aligned}
\frac{1}{\hbar^2}\bm{\sigma}\bm{P} \bm{\sigma}\bm{P}\chi =-\frac{1}{\sqrt{G}}\bar{D}_A(\sqrt{G}G^{AB} \bar{D}_B)\chi+\frac{1}{4}\bar{R}\chi,
\end{aligned}
\end{equation}
where $\bar{R}$ is the Ricci scalar.

Thus the non-relativistic equation for a spin-1/2 particle in a curved space should be described by
\begin{equation}\label{nonequ}
-\frac{\hbar^2}{2m}[\frac{1}{\sqrt{G}}\bar{D}_A(\sqrt{G}G^{AB} \bar{D}_B)-\frac{1}{4}\bar{R}]\chi=E_s\chi.
\end{equation}
where $\bar{D}_A$ is a new gauge covariant derivative with $\bar{D}_A=\partial_A+\bar{\Omega}_A$.
It is interesting that the equation Eq.~\eqref{nonequ} has the same form of the  Schr\"{o}dinger equation in an externally applied electromagnetic field.
The difference is that in the Schr\"{o}dinger equation the gauge structure is determined by the externally applied electromagnetic field with U(1) symmetry, while in Eq.~\eqref{nonequ} the gauge structure is constructed by the spin connection $\bar{\Omega}_A$. Under the SU(2) rotations of the direction $\sigma_K$, $\mathcal{T}=e^{\frac{i}{4}\epsilon^{IJK}\sigma_K \gamma_{IJ}}$, the wave function $\chi$ and the spin connection $\bar{\Omega}_A$ transforms as
\begin{equation}\label{tran}
\begin{aligned}
\chi & \rightarrow\chi^\prime=\mathcal{T}\chi,\\
\bar{\Omega}_A &\rightarrow \bar{\Omega}_A^\prime=\mathcal{T}\bar{\Omega}_A\mathcal{T}^\dagger +\mathcal{T}\partial_A \mathcal{T}^\dagger,
\end{aligned}
\end{equation}
where $\frac{1}{4}\epsilon^{IJK}\gamma_{IJ}$ plays the role of rotation angle. It is obvious that $\bar{\Omega}_A$ transforms as a gauge field in the adjoint representations of SU(2).

\section{effective equation in curved 2D space with spin connection}
The required effective equation of a spin-1/2 particle confined to a curved surface will be achieved in this section. For the achievement the thin-layer quantization approach~\cite{JENSEN1971586, PhysRevA.23.1982, Wang2016a} is suitable. According to the fundamental framework~\cite{Wang2016a}, the metric tensors $g_{ab}$ on a curved surface $S$ and $G_{AB}$ in the corresponding 3D immediate neighborhood space should be first considered. If the surface $S$ can be described by $\textbf{r}=\textbf{r}(q_1,q_2)$, the corresponding 3D space could be done by $\textbf{R}=\textbf{r}(q_1,q_2)+q_3\hat{\textbf{n}}(q_1,q_2)$. Subsequently, the associated metric tensors $g_{ab}$ and $G_{AB}$ can be defined by $g_{ab}=\partial_a \textbf{r} \cdot \partial_b \textbf{r}$ and $G_{AB}=\partial_A \textbf{R} \cdot \partial_B \textbf{R}$, which satisfy the following relation
\begin{equation}\label{metric}
\begin{aligned}
G_{ab}&=g_{ab}+[\alpha g+(\alpha g)^T]_{ab}q_3+(\alpha g \alpha^T)_{ab}(q_3)^2, \\
G_{a3}&=G_{3a}=0, \ \ G_{33}=1,
\end{aligned}
\end{equation}
where the Weingarten curvature matrix is
\begin{equation}
\alpha_{ab}=\partial_a \textbf{r}\cdot \partial_b\hat{\textbf{n}}.
\end{equation}

For the convenience of statement, we refer to the coordinate system $(q_1, q_2)$ as a surface frame (SF), and the coordinate system $(q_1,q_2,q_3)$ as an adapted frame (AF), where $q_3$ is the coordinate variable in the $\hat{\textbf{n}}$ direction. From Eq.~\eqref{metric}, the relation between the determinants $G$ and $g$ can be calculated as $G=f^2g$, where $f=1+\rm{Tr}(\alpha_{ab})q_3 +\det(\alpha_{ab})q_3^2$ named as the rescaled factor. Under the rescaled transformation, an introduced new wave function $\psi$ and the Hamiltonian in Eq.~\eqref{nonequ}~\cite{SCHUSTER2003132, Wang2018a, Wang2018b} can be expressed as
\begin{equation}\label{neww}
\psi=f^{\frac{1}{2}}\chi,
\end{equation}
and
\begin{equation}\label{newH}
{\rm{H}}^\prime=f^{\frac{1}{2}}[-\frac{\hbar^2}{2m\sqrt{G}}\bar{D}_A(\sqrt{G}G^{AB} \bar{D}_B)]f^{-\frac{1}{2}}+\frac{\hbar^2\bar{R}}{8m}.
\end{equation}
%Here the invariance of the Ricci scalar under the rescaled transformation is considered.

The introduction of the confining potential plays an essential role in the thin-layer quantization scheme. The confining potential raises the energy of normal excitations far beyond the energy scale associated with motion tangent to the surface, and entirely determine the separation of the normal and tangent dynamics. By introducing the confining potential $V_c(q_3)$, we can deduce (the calculation details are shown in Appendix~\ref{app}) the effective Hamiltonian and the normal component~\cite{Wang2018a} as below
\begin{equation}\label{effh}
\begin{split}
{\rm{H_{eff}}}+{\rm{H_N}}&=\lim_{q_3\to 0}[{\rm{H}}^\prime+V_c]\\
&=-\frac{\hbar^2}{2m}[\frac{1}{\sqrt{g}}D_a(\sqrt{g}g^{ab}D_b)-\frac{K}{2}]\\
& \quad -\frac{\hbar^2}{2m}\frac{-i}{\sqrt{g}}[\mathcal{S}^{ab} \sigma_a\partial_b+\frac{1}{2}\partial_b(\sigma_a\mathcal{S}^{ab})]\\
& \quad -\frac{\hbar^2}{2m}\partial_3^2+V_c,
\end{split}
\end{equation}
where $D_a$ denotes a gauge covariant derivative with $D_a=\partial_a+i\sigma_3 w_a$, wherein $w_a$ can be taken as a gauge potential with $w_a=\frac{1}{4}\epsilon^{ij}\omega_{aij}$, $K$ is the Gaussian curvature, and $\mathcal{S}^{ab}$ is the coupling tensor defined by $\mathcal{S}^{ab}=\epsilon^{ac}\alpha_c^b$.
As a consequence, the effective surface dynamics is
\begin{equation}\label{teq}
{\rm{H_{eff}}}\chi_t=({\rm{H_0+H_{so}}})\chi_t=E_t\chi_t,
\end{equation}
and the normal dynamics as
\begin{equation}
(-\frac{\hbar^2}{2m}\partial_3^2+V_c) \chi_r=E_r \chi_r.
\end{equation}
In Eq.~\eqref{teq} the effective Hamiltonian ${\rm{H}_{eff}}$ including two components, $\rm{H}_0$ is
\begin{equation}\label{h0}
{\rm{H}_0}=-\frac{\hbar^2}{2m}[\frac{1}{\sqrt{g}}D_a(\sqrt{g}g^{ab}D_b)-\frac{K}{2}],
\end{equation}
and  $\rm{H_{so}}$ describing the SOI on the curved surface $S$ in the following form
\begin{equation}\label{hso}
{\rm{H_{so}}}=-\frac{\hbar^2}{2m}\frac{-i}{\sqrt{g}}[\mathcal{S}^{ab} \sigma_a\partial_b+\frac{1}{2}\partial_b(\sigma_a\mathcal{S}^{ab})].
\end{equation}
Notice that the geometric potential $-\frac{\hbar^2}{2m}(-\frac{K}{2})$ in Eq.~\eqref{h0} is striking different from the well-known form $-\frac{\hbar^2}{2m}(M^2-K)$~\cite{PhysRevA.23.1982}. The well-known geometric potential is completely determined by the reduced communication relation~\cite{Wang2017}, while the appearance of the spin connection $\bar{\Omega}_A$ in the gauge covariant derivative $\bar{D}_A$ leads to the term $-\frac{\hbar^2}{2m}\bar{\Omega}_a\bar{\Omega}^a$ appearing in Eq.~\eqref{h0} as an additional term $-\frac{\hbar^2}{2m}(\frac{K}{2}-M^2)$. In other words, the presence of the spin degree of freedom can influence the geometric effects of the curved surface~\cite{PhysRevA.90.042117}. Besides, the term $\frac{\hbar^2 \bar{R}}{8m}$ has no contribution to the scalar potential as Ricci scalar vanishes in the thin-layer procedure.

Furthermore, it is worthwhile to notice that the gauge potential $w_a$ and the effective SOI $\rm{H_{so}}$ are present in the effective Hamiltonian $\rm{H_{eff}}$ due to the appearance of the spin connection. Mathematically, the spin connection is determined by the derivative operators and the rotation transformation of the spin orientation in adjacent different local positions. Therefore, the geometry of the curved surface can be used to deform the forms of the pseudo-magnetic field defined by the gauge potential $w_a$ and the effective SOI. The two geometric effects will be given further discussions.

%From the effective tangential equation, we find the continuity equation is
%\begin{equation}
%\begin{aligned}
%\partial_t(\psi^*\psi)=&\frac{1}{2m}\bm{\nabla}[\psi^*(\bm{P}-i\hbar\sqrt{g}\sigma_3 \bm{w}-i\hbar\bm{U}\bm{\sigma})\psi \\
%&+\psi (\bm{P}-i\hbar\sigma_3 \bm{w}-i\hbar\bm{U}\bm{\sigma})^* \psi^* ],
%\end{aligned}
%\end{equation}
%where $\bm{U}=\frac{i}{2}\epsilon^{bc}\alpha_b^a$ is a tensor,
%so we obtain the conserved charge $\rho=\psi^*\psi$, and the conserved current
%\begin{equation}
%\bm{j}=Re(\psi^* \hat{\bm{v}}\psi),
%\end{equation}
%where the velocity operator $\hat{v}$ can be defined as
%\begin{equation}
%\hat{v}^a=-\frac{i\hbar \sqrt{g}}{m}(\partial^a-\sigma_3 w^a - \frac{i}{2\sqrt{g}}\epsilon^{bc}\sigma_c \alpha_b^a).
%\end{equation}
%The expression of the velocity operator displays that particles with different spin orientation may be distinguished based on their motions. In fact this kind of phenomenon can be attributed to the pseudo-magnetic field and effective SOI.
%Eq.~\eqref{teq} is one of the main finding of this paper. It shows that even without external electromagnetic field, the effective SOI and pseudo-magnetic field as purely geometric effects appear for a spin-1/2 particle in a curved surface. The pseudo-magnetic field and effective SOI are in fact the results of symmetry breaking from the spin connection. Because of the confining potential, the freedom of the gauge choice associated with the normal direction vanishes.
%In the following we discuss the two effects respectively.

\subsection{Pseudo-magnetic field}
It is easy to prove that the effective dynamics Eq.~\eqref{teq} still possesses the invariance of a rotation transformation, $\mathcal{T}_3=e^{i\sigma_3\theta}$ with $\theta=\frac{1}{4}\epsilon^{ij}\gamma_{ij}$. Under the rotation transformation $\mathcal{T}_3$, $w_a$ transforms as a gauge potential, that is
\begin{equation}\label{gaug}
w_a \rightarrow w_a^\prime=\mathcal{T}_3w_a\mathcal{T}_3^\dagger+\mathcal{T}_3\partial_a\mathcal{T}_3^\dagger.
\end{equation}
In terms of the gauge potential $\bm{w}$, the pseudo-magnetic field can be defined by
\begin{equation}\label{pmagnetic}
\mathcal{B}=-\frac{\hbar}{e}\bm{\nabla}_{(2D)} \times \bm{w}=\frac{\hbar K}{2e},
\end{equation}
where $\bm{\nabla}_{2D} \times \bm{w}=-\frac{K}{2}$ and the curl operator is defined in 2D spaces. In the calculation we have utilized the formula $R_{1212}/g=-K$, where $R_{abcd}$ is the Riemann curvature tensor. We notice that in Ref.\cite{PhysRevB.92.245110} the pseudo-magnetic field calculated for curved graphene systems is twice of this result. The reason is that the spin connection used in Ref.\cite{PhysRevB.92.245110} is twice of ours. According to the pseudo-magnetic field Eq.~\eqref{pmagnetic}, we estimate the value for a bubble with radius $r \sim 1\text{nm}$ as $\mathcal{B} \sim 328\text{T}$, which is also in agreement with the result given experimentally in graphene systems~\cite{Levy544}.
%Comparing with the gauge field in low energy effective Dirac equation for graphene with deformation, $w_a$ in Eq.

Obviously, in Eq.~\eqref{pmagnetic} the gauge field is totally determined by the Gaussian curvature. As a consequence, we can provide a required gauge potential for spin-1/2 particles confined to a curved surface by designing the geometry of the surface. For example, in an orthogonal curvilinear coordinate system $(q_1,q_2)$, we can choose $w_1=-\frac{1}{2}\int \sqrt{g} K dq_2$ and $w_2=0$.

It's well known that the external magnetic field breaks the time reversal symmetry, however, this doesn't happen for the pseudo-magnetic field. This conclusion can be tested by obtaining the commutator $[T,H_0]=0$, where the time-reversal operator $T$ is choosen as $T=i\sigma_y C$ with $C$ the complex-conjugation operator. A physical interpretation is that the pseudo-magnetic field couples with the matrix $\sigma_3$ rather than a scalar constant, hence under time inverse, the pseudo-magnetic field is reversed spatially or changes it's sign accordingly, keeping the total system invariant. The largeness and time reversal symmetry of pseudo-magnetic field may help to manufacture materials with topological properties. On the one hand, the application of quantum Hall effect requires large magnetic field, on the other hand, the time reversal symmetry protects the robustness of the quantum spin Hall edge state~\cite{Qi2010The}. Hence by bending the 2D materials properly, topological states may emerge.

Another interesting property of the pseudo-magnetic field is about the topology in a real space. According to Gauss-Bonnet theorem, for a surface $S$ without a boundary, we have
\begin{equation}
\frac{1}{2\pi}\int_S K dA=2(1-g),
\end{equation}
where $g$ is the genus of the surface. Straightforwardly, the pseudo-magnetic flux is
\begin{equation}
\Phi=\int_S \mathcal{B} dA=2(1-g)\Phi_0,
\end{equation}
where $\Phi_0=\frac{h}{2e}$ is the magnetic flux quantum. This result shows that the pseudo-magnetic flux for a closed surface is a topological invariant, which is independent of it's geometric details. For example, the pseudo-magnetic flux for a sphere and a torus is $2\Phi_0$ and 0, respectively, no matter how their size and shape change continuously.

Particles with opposite charge will be separated by Lorentz force when they are moving in a magnetic field, which is known as Hall effect. In the presence of the pseudo-magnetic field, the spin-1/2 particles with different spin orientations in the normal direction should also be separated by a Lorentz-like force. The phenomenon is the spin Hall effect, which will be found in a bent cylindrical surface.

\subsection{Effective SOI}\label{esoi}

In Eq.~\eqref{hso}, the second term is not negligible as it makes sure that the Hamiltonian is Hermitian. The expression is similar to that of SOI caused by an electric field $\bm{E}$~\cite{16322}, namely
\begin{equation}
\tilde{H}_{so}=\frac{i\hbar^2 e}{4m^2c^2}[\bm{\sigma}(\bm{E}\times \bm{\nabla})-\bm{\sigma}(\bm{\nabla}\times \bm{E})/2],
\end{equation}
where the second term provides the hermiticity of the Hamiltonian. However, we can't simply refer this effective spin-orbit interaction as the effect induced by a ``pseudo-electric field",
since in general the coupling tensor $\mathcal{S}^{ab}$ is anisotropic. Specifically, when $\alpha_1^1=\alpha_2^2$, the Hamiltonian has the form of linear Rashba spin-orbit interaction, and if $\alpha_1^2=\alpha_2^1$, the Hamiltonian contain the form of linear Dresselhaus spin-orbit interaction. In the case where the surface has the symmetry $\alpha_1^1=\alpha_2^2$, we can define a ``pseudo-electric field" which is oriented to the normal direction, that is
\begin{equation}
\mathcal{E}=\frac{2mc^2 \alpha_1^1}{e}.
\end{equation}

As the situation in the planar 2DEG, this ``pseudo-electric field" results in spin Hall effect. We have mention that the pseudo-magnetic field also leads to spin Hall effect, thus it is worthwhile comparing the two effective interactions. It is convenient to follow the previous research and reexpress the the SOI as a non-Abelian gauge field, which is exactly $(A_{so})_a$ in Eq.~\eqref{scs}. Using Eq.~\eqref{psig} and ~\eqref{chris}, we obtain the curl
\begin{equation}
\begin{aligned}
\bm{\nabla}_{(2D)}\times (\bm{A}_{so})=&-\frac{K}{2}\sigma_3+F^a(q_1,q_2) \sigma_a \\
=&\sigma_3\bm{\nabla}_{(2D)}\times \bm{w}+F^a(q_1,q_2) \sigma_a,
\end{aligned}
\end{equation}
where $F^a (q_1,q_2)$ is a vector function of the coordinates.
Interestingly, the part of this curl that is coupled with $\sigma_3$, is equal to the curl of the gauge field $\bm{w}$. For the spin orientations normal to the surface, the effective SOI seems to provide an effective field which is identical with the pseudo-magnetic field.
That is to say the effective SOI and the pseudo-magnetic field may contribute exactly the same forces for the spin Hall effect. We will demonstrate this conjecture on a bent cylindrical surface. Besides, we have to emphasize that, although the pseudo-magnetic field and the effective SOI show common effects on the normal spin component, they are not equivalent for the dynamics of spin components tangential to the surface.

To estimate when the effective SOI can be ignored, we need compare the coupling strengths between the effective SOI and intrinsic SOI of specific materials.
It is found that if the two coupling strengths are commensurable, the curvature radius should be
\begin{equation}\label{soir}
r \sim \frac{\hbar^2}{2m\tilde{\alpha}} \approx \frac{3.79 \times 10^{-20}\rm{eV} \cdot m^2}{\zeta \tilde{\alpha}},
\end{equation}
where $\zeta=m/m_e$ with $m$ the effective mass and $m_e$ the electron rest mass. For instance, the intrinsic SOI coupling constant of InGaAs~\cite{PhysRevB.69.035302} is $\tilde{\alpha}=(3 \sim 4 ) \times 10^{-11}$eV$\cdot$m, and ratio $\zeta=0.041$, then $r = 23 \sim 31$nm. This result show that the effective SOI can not be neglected as the curvature radius reaches nanoscale.

%From the expression of the current, the spin connection in $AF$ is decomposed into a 2D gauge field coupling with $\sigma_3$ and a term coupling with the tangential components of Pauli matrix.

\section{Straight and bent cylinders}
In this section, using the previous results we will investigate two simple examples, a straight cylindrical surface and a bent one.
\subsection{Straight cylinders}
\begin{figure}
  \centering
  \includegraphics[width=0.4\textwidth]{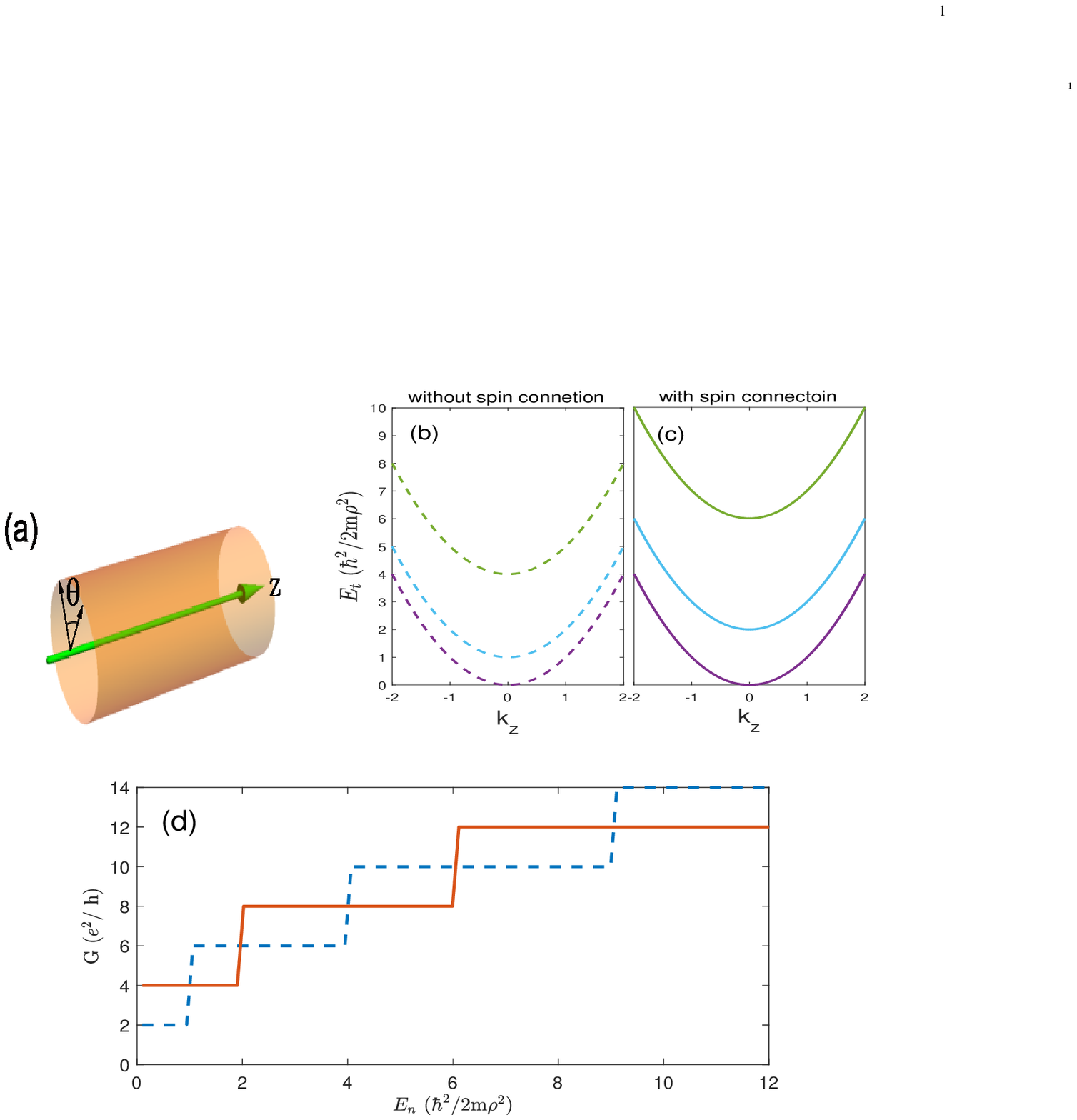}\\
  \caption{(Color online) (a) Illustration of a cylinder. (b) Energy dispersion diagram without the consideration of the spin connection. (c) Energy dispersion diagram with the consideration of the spin connection. (d) Conductance of a cylinder with (solid line) and without (dash line) the consideration of the spin connection at zero temperature.}\label{cylin}
\end{figure}

In a cylindrical coordinate system $(\rho,\theta,z)$, as shown in Fig.~\ref{cylin}(a), the effective equation for a straight cylinder is
\begin{equation}\label{cyli}
i\hbar \partial_t \psi_t=-\frac{\hbar^2}{2m}[\partial_z^2+\frac{1}{\rho^2}(\partial_\theta^2+i\sigma_z \partial_\theta)]\psi_t,
\end{equation}
where the spin-orbit coupling term $-i\frac{\hbar^2}{2m\rho^2}\sigma_z \partial_\theta$ is just contributed by the spin connection. In this simple geometry, the pseudo-magnetic field disappears because of the vanishing of the Gaussian curvature, but the effective SOI still exist, showing the difference between the pseudo-magnetic field and the effective SOI.

Neglecting the spin-orbit coupling term, as what we used to do, in $\theta$ direction, the corresponding energy spectrum and wavefunctions are $\tilde{E}_n=\frac{n^2\hbar^2}{2m\rho^2}$ and $|n,\pm\rangle=\bm{u}_\pm e^{in\theta}$, respectively, where $\bm{u}_+=(1,0)^T$ and $\bm{u}_-=(0,1)^T$ are the eigenstates of the Pauli matrix $\sigma_z$ and the magnetic quantum number $n=0,\pm1,\pm2,\cdots$, as shown in Fig.~\ref{cylin}(b). In this case only the ground state energy is 2-fold degenerate, and the other excited energies are 4-fold degenerate. While if we take into account the spin connection contribution, the energy levels become $E_n=\frac{\hbar^2}{2m\rho^2}(n^2\pm n)$ [Fig.~\ref{cylin}(c)]. With the shift of energy levels, the degeneration becomes different. As the mathematical relation $(n\pm1)^2\mp(n\pm1)=n^2\pm n$, all energy levels are 4-fold degenerate, including the ground state energy, whose eigenstates are expressed as $|0,+\rangle, |0,-\rangle, |1, -\rangle, |-1,+ \rangle$. In fact, the corresponding Hamiltonian in Eq.~\eqref{cyli} can be written as $H=\frac{p_z^2}{2m}+\frac{1}{2m\rho^2}\hat{J}_z^2-\frac{\hbar^2}{8m\rho^2}$, where the total angular momentum $\hat{J}_z=\hat{L}_z+\hat{s}_z=-i\hbar \partial_\theta+\frac{\hbar}{2}\sigma_z$, and the well known geometric potential $-\frac{\hbar^2}{8m\rho^2}$ is recovered. Hence we ought to apply new quantum number $j_\pm=n\pm 1/2$ and eigenstate notation $|j_\pm,\pm\rangle$ to describe the system, which acts as replacing the orbital angular momentum with the total angular momentum in the Hamiltonian without spin connection contribution. Indeed, including the spin connection makes the model more symmetric and physically sound. Experimentally, the conductance in a cylindrical 2DEG may manifest the existence of the effective SOI. As shown in Fig.~\ref{cylin}(d), the steplike structures of the conductance show apparent difference between considering and neglecting the effective SOI at zero temperature.

\subsection{Bent cylinders}
\begin{figure}
  \centering
  \includegraphics[width=0.4\textwidth]{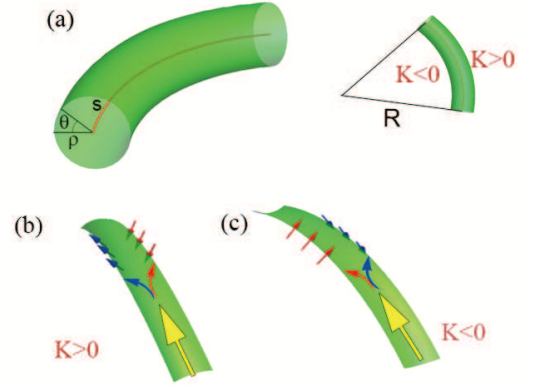}\\
  \caption{(Color online) (a)A bent cylindrical surface. The Gaussian curvature $K>0$ at the outer points, and $K<0$ at inner points. (b) The outer part of the bent cylindrical surface, $\theta\in [-\theta_0,\theta_0]$, where $\theta_0$ is small. The yellow arrow denotes the incident charge current, and the red and blue arrows denote the currents with spin orientation pointing inwards and outwards, respectively. (c) The inner part of the bent cylindrical surface, $\theta\in [\pi-\theta_0,\pi+\theta_0]$.}\label{bent}
\end{figure}

To investigate the combined effect of both the pseudo-magnetic field and the effective SOI, we consider a bent cylindrical surface [Fig.~\ref{bent}(a)], which can be described in the curvilinear coordinate system $(\theta,s)$, where $\theta$ is the angle around the cylinder's axis, and $s$ the arclength of the axis. The corresponding Hamiltonian is written as
\begin{equation}
\begin{aligned}
H_0=&-\frac{\hbar^2}{2m\rho^2}\frac{R}{R+\rho\cos\theta}\partial_\theta
(\frac{R+\rho\cos\theta}{R}\partial_\theta ) \\
&-\frac{\hbar^2}{2m}\frac{R^2}{(R+\rho\cos\theta)^2}[\partial_s-\frac{i\sigma_3}{2R}\sin\theta][\partial_s- \frac{i\sigma_3}{2R}\sin\theta] \\
& +\frac{\hbar^2}{4m}\frac{\cos \theta}{\rho(R+\rho\cos \theta)}
\end{aligned}
\end{equation}
and
\begin{equation}
\begin{aligned}
H_{so}=\frac{i\hbar^2}{2m}\frac{R}{\rho(R+\rho \cos\theta)}(\frac{\cos\theta}{R+\rho\cos\theta}\sigma_\theta \partial_s-\frac{1}{\rho}\sigma_s \partial_\theta),
\end{aligned}
\end{equation}
where $\rho$ is the radius of the cylinder, and $R$ the curvature radius of the axis.
%The sphere is also a simple curved geometry to investigate the curvature induced effect, unlike the cylinder, it owns a constant Gaussian curvature, which allows us to study the pseudo-magnetic field. In the spherical coordinate system $(r,\theta,\varphi)$, we write the corresponding effective Hamiltonian as
%\begin{equation}
%\begin{aligned}
%H_0=&-\frac{\hbar^2}{2mr^2}\frac{1}{\sin\theta}(\partial_\theta+\frac{i}{2}\sigma_r \varphi \sin\theta)[\sin\theta(\partial_\theta+\frac{i}{2}\sigma_r \varphi \sin\theta) ] \\
%&-\frac{\hbar^2}{2mr^2\sin^2 \theta}\partial^2_\varphi
%\end{aligned}
%\end{equation}
%and
%\begin{equation}
%\begin{aligned}
%H_{so}=&\frac{i\hbar^2}{2mr^3 \sin\theta}[\sigma_\theta\partial_\varphi-\sigma_\varphi\partial_\theta].
%\end{aligned}
%\end{equation}
Using the Heisenberg equation of motion, we obtain
\begin{equation}
i\hbar\dot{\theta}=[\theta,H_0+H_{so}],
\end{equation}
where
%\begin{equation}
%i\hbar\dot{\varphi}=[\varphi,H]=\frac{\hbar^2}{mr^2\sin^2\theta}\partial_\varphi-\frac{i\hbar^2}{2mr^3 \sin\theta}\sigma_\theta,
%\end{equation}
%then
%\begin{equation}
%\ddot{\varphi}=
%\end{equation}
\begin{equation}
[\theta,H_0]=\frac{\hbar^2}{m\rho^2}[\partial_\theta-\frac{\rho\sin\theta}{2(R+\rho\cos \theta)}],
\end{equation}
and
\begin{equation}
[\theta,H_{so}]=\frac{i\hbar^2}{2m}\frac{R}{\rho^2(R+\rho\cos\theta)}\sigma_s.
\end{equation}

To reveal the effect of the pseudo-magnetic field, we give the boundary condition that $\theta \in [-\theta_0,\theta_0]$ and $R \gg \rho$, where $\theta_0$ is small so that $\sin\theta \approx 0$, $\cos\theta \approx 1$, as shown in Fig.~\ref{bent}(b). In addition, a charge current is injected along the axis $s$ with the wave length $\lambda<\rho$.
In this condition, the force in $\theta$ direction associate with $\sigma_3$ is dominantly determined by the current. We prefer calculating the force generated from the pseudo-magnetic field and the effective SOI separately. For the pseudo-magnetic field,
\begin{equation}
\ddot{\theta}_{(pm)}=\frac{1}{i\hbar}[\dot{\theta},H_0] \approx \frac{\hbar p_s R\cos\theta \sigma_3}{2m^2\rho^2(R+\rho\cos\theta)^2},
\end{equation}
where $p_s=-i\hbar\partial_s$ is the momentum operator along the axis $s$. For the effective SOI, we find
\begin{equation}
\ddot{\theta}_{(so)}=\frac{1}{i\hbar}[\dot{\theta},H_{so}] \approx \frac{\hbar p_s R\cos\theta \sigma_3}{2m^2\rho^2(R+\rho\cos\theta)^2}.
\end{equation}
Here the relation $\partial_\theta(\sigma_\theta)=\rho\sigma_3$ is utilized.
As predicted in Section~\ref{esoi}, the forces from the pseudo-magnetic field and the effective SOI are equal. Hence the total force
\begin{equation}
F_\theta=m\rho^2(\ddot{\theta}_{(pm)}+\ddot{\theta}_{(so)}) \approx 2\sigma_3 e \mathcal{B}v_s ,
\end{equation}
where $\mathcal{B}=\frac{\hbar}{2e}\frac{\cos \theta}{\rho(R+\rho\cos \theta)}$ and $v_s=p_s/m$.
It is found that this geometry induced force is in a form similar to the Lorentz force, leading to spin Hall effect.
In the representation of $\sigma_3$, spin-up particles and spin-down particles are acted on by the force in opposite direction, respectively, or equivalently, the particles in the two different spin states feel the pseudo-magnetic field in different directions, that is parallel and antiparallel to the normal direction. The same phenomenon appears when $\theta\in [\pi-\theta_0,\pi+\theta_0]$ (see Fig.~\ref{bent}(c)), only here the Gaussian curvature is negative, which induces opposite forces for the inward and outward spin orientations compared with the case where $\theta \in [-\theta_0,\theta_0]$.

\section{Conclusion}
In this paper, we started from Dirac equation in curved space and performed the non-relativistic limit and thin-layer procedure to obtain the effective dynamics for a spin-1/2 particle constrained to an arbitrary curved surface. We have shown that an effective SOI, a 2D gauge field and a scalar potential appear in the effective equation as contribution of the spin connection. The effective SOI and the gauge field are associated with the spin orientation parallel and orthogonal to the surface, respectively. Furthermore, we have found the pseudo-magnetic field generated by the gauge field is proportional to the Gaussian curvature, which makes the corresponding flux for a closed surface is topological invariant. The effective spin-orbit coupling strength is determined by the Weingarten curvature tensor, being comparable with the intrinsic SOI strength of semi-conducting materials when the curvature radius is in nanoscale.
 %In general, even without the electric and magnetic field, the effective SOI and the pseudo-magnetic field could be induced purely by the curvature.

To manifest the effect of the pseudo-magnetic field and the effective spin-orbit interaction, we have proposed two different geometries, namely a straight and a bent cylinder. In the straight cylinder, energy level shifts due to the effective SOI have been displayed, and accordingly we suggest a conductance experiment to demonstrate the existence of this interaction.%  the total angular momentum appears naturally to replace the orbit angular momentum.
In the bent cylinder, Lorentz-like force from the pseudo-magnetic field and spin-orbit force from the effective SOI have been found equal and lead to spin Hall effect.

%Thus  both the electric and magnetic field induced effects may be reconstructed in correspondingly curved geometries as a result of the spin connection.

Our model gives a lucid picture of a spin-1/2 particle moving in a 2D curved space with confinement and is well suited to the study of 2DEG systems with curved features. It shows that by bending the 2D semi-conducting materials in nanoscale, the effective SOI and pseudo-magnetic field from the spin connection could be locally enormous, which are easier to control than the external electric and magnetic field.% The two interactions could significantly change the energy band structure, endowing the materials with novel properties.
This implies new possibility in constructing materials with topological properties.
%We look forward that experimental measurements of the conductance and Landau levels could demonstrate the existence of the effective SOI and pseudo-magnetic field in curved 2DEG systems.

\acknowledgments

This work is supported in part by the National Natural Science Foundation of China (under Grants No. 11690030, No. 11475085, No. 11535005 and No. 61425018) and National Major state Basic Research and Development of China (2016YFE0129300).
Y.-L. W. was funded by the Natural Science Foundation of Shandong Province of China (Grant No.
ZR2017MA010). We would like to thank Fan Wang for enlightening discussions.

\appendix
\section{Separation of the normal and tangential dynamics}\label{app}
%How are you?
Because of the spin connection and Ricci scalar contained Eq.~\eqref{newH}, the process of taking limit $q_3 \rightarrow 0$ is more complex than the case of scalar Schr\"{o}dinger equation. We would do some preparation for this purpose.
In $AF$, the vierbein components $E_A^I$ are determined by the choice of the flat space coordinates $x^I$. For convenience, we will choose $x^I$ as the local-flat space coordinates which make the components have the form~\cite{MARANER20082044,BRANDT20163036}
\begin{equation}
E_A^I=\left(
\begin{array}{ccc}
 E_a^i &0 \\
0& 1
\end{array}
\right).
\end{equation}
The inverse of $E_A^I$ is then
\begin{equation}
E_I^A=\left(
\begin{array}{ccc}
 E_i^a &0 \\
0& 1
\end{array}
\right).
\end{equation}

In $SF$, we can also define vielbeins $e_a^i$ to satisfy the relation $g_{ab}=e_a^i e_b^j \delta_{ij}$. From Eq.\eqref{metric}, it is easy to find the relation between $E_a^i$ and $e_a^i$, that is
\begin{equation}\label{vier}
E_a^i=e_a^i+q_3 \alpha_a^b e_b^i.
\end{equation}
For the inverse ones, up to the first-order in $q_3$, it is given by
\begin{equation}\label{invi}
E_i^a=e_i^a- q_3\alpha_b^a e_i^b+O[(q_3)^2].
\end{equation}
In addition to the difference between $E_a^i$ and $e_a^i$, the Christoffel symbols and spin connections in $AF$ and $SF$ also possess different forms. To distinguish these geometric quantities, we mark the ones in $AF$ with a bar. For example, in $AF$, $\bar{\Gamma}_{ab}^c=\frac{1}{2}G^{cd}(\partial_b G_{da}+\partial_a G_{db}-\partial_d G_{ab})$, while in $SF$, $\Gamma_{ab}^c=\frac{1}{2}g^{cd}(\partial_b g_{da}+\partial_a g_{db}-\partial_d g_{ab})$.

Using Eq.~\eqref{metric}, we can find
\begin{equation}\label{chris}
\begin{aligned}
\bar{\Gamma}_{ab}^c&=\Gamma_{ab}^c+O(q_3), \\
\bar{\Gamma}_{3a}^b&=\bar{\Gamma}_{a3}^b=\alpha_a^b-q_3(\alpha g \alpha^T)_a^b+O[(q_3)^2], \\
\bar{\Gamma}_{ab}^3&=-\alpha_{ab}-(\alpha g \alpha^T)_{ab} q_3, \\
\bar{\Gamma}_{33}^A&=\bar{\Gamma}_{3A}^3=\bar{\Gamma}_{A3}^3=0.
\end{aligned}
\end{equation}
Then, by using Eq.~\eqref{omeg},~\eqref{vier},~\eqref{invi}and Eq.~\eqref{chris}, we calculate the spin connection in $AF$ and obtain
\begin{equation}\label{scs}
\begin{aligned}
\bar{\Omega}_a&=\Omega_a+i(A_{so})_a+O(q_3), \\
\bar{\Omega}_3&=O[(q_3)^2],
\end{aligned}
\end{equation}
where $\Omega_a=i\sigma_3w_a=\frac{i}{4}\sigma_3\epsilon^{ij} \omega_{aij}$, and $(A_{so})_a=\frac{1}{2\sqrt{g}}\epsilon^{cb}\sigma_b\alpha_{ac}$, which can be viewed as spin connections in $SF$ and non-Abelian spin-orbit gauge field, respectively. Eq.~\eqref{scs} is the most important step in the separation of tangential and normal dynamics.
It is clear that the spin connection is decomposed into two parts: one is associated with the normal component of Pauli matrices, and the other one couples with the tangential components.

Besides, we have to deal with the scalar potential $\frac{\bar{R}}{8m}$ carefully. The Ricci scalar can be calculated according to the formula
\begin{equation}
\bar{R}=G^{AB}\bar{R}_{AB}=G^{ab}\bar{R}_{ab}+\bar{R}_{33},
\end{equation}
where $\bar{R}_{AB}$ is the Ricci tensor. In a pure two-dimensional space, or $SF$, the Ricci scalar $R=g^{ab}R_{ab}$ is exactly twice the Gaussian curvature, however, in $AF$, this is not right even at $q_3=0$. We will show this below.

First, we find
\begin{equation}
R_{33}=-\Gamma_{3a}^b\Gamma_{b3}^{a}-\partial_3\Gamma_{3a}^a=O(q_3),
\end{equation}
then the term
\begin{equation}
\begin{aligned}
G^{ab}\bar{R}_{ab}&=g^{ab}R_{ab}+ \\
&g^{ab}(\partial_3 \Gamma_{ba}^3-\Gamma_{b3}^c \Gamma_{ca}^3-\Gamma_{bc}^3 \Gamma_{3a}^c+\Gamma_{ab}^3\Gamma_{3c}^c)+O(q_3) \\
&=2K-2K+O(q_3)=O(q_3).
\end{aligned}
\end{equation}
Therefore in the limit $q_3\rightarrow 0$ the scalar potential $\frac{\hbar^2 \bar{R}}{8m}$ in Eq.~\eqref{nonequ} vanishes instead of being $\frac{\hbar^2 K}{4m}$.

So far, we are ready to perform the limit $q_3 \rightarrow 0$ in the Eq.~\eqref{newH} and separate the equation into tangential and normal components. Substituting the results above, we achieve Eq.~\eqref{effh}.
%\section{The second}
%All is well!

%\section*{appendix: Structure of the quark propagator}\label{appA}
%Nice.

\bibliographystyle{apsrev4-1}
\bibliography{ref}

%\begin{thebibliography}{99}
%\bibitem{a14}  M. Stephanov, K. Rajagopal and E. Shuryak, Phys. Rev. Lett. 81, 4816 (1998).
%\bibitem{c4} H. S. Zong, L. Chang, F. Y. Hou, W. M. Sun and Y. X. Liu, Phys. Rev. C 71, 015205 (2005);
%             F. Y. Hou, L. Chang, W. M. Sun, H. S. Zong, and Y. X. Liu, Phys. Rev. C 72, 034901 (2005).
%\end{thebibliography}

\end{document}